\documentclass{amsproc}

\usepackage{graphicx}
\usepackage{latexsym}

\newtheorem{theorem}{Theorem}[section]

\theoremstyle{definition}

\theoremstyle{remark}

\numberwithin{equation}{section}

\newcommand {\ket}[1]     {|{#1}\rangle}

\newcommand {\setof}[1]   {\left\{{#1}\right\}}
\newcommand {\sqmatrix}[4] {\left( \begin{array}{cc}
                                    {#1} & {#2} \\ 
				    {#3} & {#4} 
				    \end{array}
		            \right)}

\newcommand {\onesqrt}   {{1 \over \sqrt 2}}

\newcommand {\oneof}[1]  {{1 \over {#1}}}
\newcommand {\epr}       {{\onesqrt(\ket{00}+\ket{11})}}

\begin{document}

\title{Generalized GHZ States and Distributed Quantum Computing}

\author{Anocha Yimsiriwattana}
\address{Department of Computer Science and Electrical Engineering, University
of Maryland, Baltimore County, Baltimore, Maryland 21250}
\email{ayimsi1@umbc.edu}


\author{Samuel J. Lomonaco Jr.}
\email{lomonaco@umbc.edu}
\thanks{This work is supported by DARPA}

\date{January 26, 2004.}



\begin{abstract}

A key problem in quantum computing is finding a viable technological path
toward the creation of a scalable quantum computer. One possible approach
toward solving part of this problem is distributed computing, which provides an
effective way of utilizing a network of limited capacity quantum computers.  

In this paper, we present two primitive operations, cat-entangler and
cat-disentangler, which in turn can be used to implement non-local operations,
e.g. non-local CNOT and quantum teleportation. We also show how to establish an
entangled pair, and use entangled pairs to efficiently create a generalized GHZ
state. Furthermore, we present procedures which allow us to reuse channel
qubits in a sequence of non-local operations.  

These non-local operations work on the principle that a cat-like state, created
by cat-entangler, can be used to distribute a control qubit among multiple
computers. Using this principle, we show how to efficiently implement non-local
control operations in many situation, including a parallel implementation of a
certain kind of unitary transformation. Finally, as an example, we present a
distributed version of the quantum Fourier transform. 

\end{abstract}

\maketitle
\section{\label{intro}Introduction}

Distributed computing provides an effective means of utilizing a network of
limited capacity quantum computers. By connecting a network of limited capacity
quantum computers via classical and quantum channels, a group of small quantum
computers can simulate a quantum computer with a large number of qubits. This
approach is useful for the development of quantum computers because the
earliest useful quantum computers will most likely hold only a small number of
qubits. This constraint limits the usage of such quantum computers to small
problems.  

We propose that distributed quantum computing (DQC) is a possible solution to
this problem. Furthermore, even if one could construct a large quantum
computer, the distributed computing model can still provide an effective means
of increasing computational power.

By a distributed quantum computer, we mean a network of small quantum
computers, connected by classical and quantum channels. Each quantum computer
(or node) has a register that can hold only a limited number of qubits.  Each
node also possesses a small number of channel qubits which can be sent back
and forth over the network. A qubit in a register can freely interact with any
other qubit in the same register. It also can freely interact with channel
qubits that are in the same node. To interact with qubits on a remote
computer, the qubits have to be teleported or physically transported to the
remote computer, or have to interact via non-local operations.

Indeed, distributed quantum computing can simply be implemented by teleporting
or physically transporting qubits back and forth. A more efficient
implementation of DQC has been proposed by Eisert et al~\cite{non-local} using
a non-local CNOT gate. Since the control NOT gate (CNOT) together with all
one-qubit gates is universal~\cite{factoring}, a distributed implementation of
any unitary transformation reduces to the implementation of non-local CNOT
gates.  Eisert et al also prove that one shared entangled pair (ebit) and two
classical bits (cbits) are necessary and sufficient to implement a non-local
CNOT gate.

In this paper, we present two primitive operations, cat-entangler and
cat-disentangler, which in turn can be used to implement non-local operations,
e.g. a non-local CNOT and quantum teleportation protocol. The cat-entangler and
cat-disentangler can be implemented using only local operation and classical
communication (LOCC), assuming that an entanglement has already been
established. We show how to establish an entangled pair between two nodes, and
use entangled pairs to efficiently create a generalized GHZ state. Furthermore,
we present procedures which allow us to reuse channel qubits in a sequence of
non-local operations.

To implement a non-local CNOT gate, first an entangled pair must be
established between two computers. Then the cat-entangler transforms a control
qubit $\alpha \ket{0} + \beta \ket{1}$ and an entangled pair $\epr$ into the
state $\alpha \ket{00} + \beta \ket{11}$, called a ``cat-like'' state. This
state allows two computers to share the control qubit.  As a result, each
computer now can use a qubit shared within the cat-like state as a local
control qubit. After completion of the control operation, cat-disentangler is
then applied to disentangle and restore the control qubit from the cat-like
state.  Finally, the channel qubits are then be reset so that the entangled
pair can be re-established.

To teleport an unknown qubit to a target qubit, we begin by establishing an
entangled pair between two computers. Then we apply the cat-entangler operation
to create a cat-like state from an unknown qubit and the entangled pair. After
that, we apply a slightly modify cat-disentangler operation to disentangle and
restore the unknown qubit from the cat-like state into the target qubit.
Finally, we reset the channel qubits. In other words, quantum teleportation can
be considered as a composition of the cat-entangler operation followed by the
cat-disentangler operation.

The cat-entangler and cat-disentangler operations can be extended to a
multi-party environment by replacing the entangled pair with a generalized
Greenberger-Horne-Zeilinger (GHZ) state (also called a cat state) expressed as
$\onesqrt( \ket{00 \ldots 0} + \ket{11 \ldots 1})$. The state $\alpha \ket{00
\ldots 0} + \beta \ket{11 \ldots 1}$ (also called a cat-like state) can be
created using only LOCC.  A cat-like state can be used to share a control qubit
between multiple computers, allowing each computer to use a qubit shared within
the cat-like state as a local control line.  In many cases, this idea leads to
an efficient implementation of multi-party control gates. In addition, a
parallel implementation of some unitary transformations can also be realized.

Before performing any non-local operation, an entanglement between computers
must be established. Moreover, the entanglement has to be refreshed after its
use.  Brennen, Song, and Williams address this issue in a lattice model quantum
computer using entanglement swapping~\cite{entswap}, which can be used to
establish and refresh entanglement between two qubits. The multiple
entanglement swapping~\cite{multi-entswap} can create a generalized GHZ state,
which is used by multiple computers.  

We address these same issues for the quantum network model by showing how to
establish two entangled pairs by sending two qubits, one in each direction.
Asymptotically, this is equivalent to establishing one entangled pair at the
cost of sending one qubit. Furthermore, we show how to convert a number of
entangled pairs into a generalized GHZ state, which in turn is used to
distribute control over multiple computers. 

We also address refreshing entanglement by observing that measurements, made
during the primitive operations, provide crucial information for resetting
channel qubits to $\ket{0}$. Hence, channel qubits can be re-entangled with
other channel qubits at a later time.

The idea of using a cat-like state to distribute control qubits is discussed in
section~\ref{sec:dist-ctrl}. Next, the cat-entangler and cat-disentangler are
presented. After that, we use these operations to construct non-local CNOT and
teleportation operations.  Section \ref{sec:ctrl-gates} discusses various
constructions of non-local control gates in different situations. Issues
related to establishing and refreshing entanglement are addressed in
section~\ref{sec:impt-em} via constructing entangling gates.  Finally, an
example of a distributed version of the quantum Fourier transform is presented
in section~\ref{sec:fourier}.

\section*{\label{sec:notation} Notation}
We adopt the notation found in~\cite{elem-gate}. For any one-qubit unitary
matrix \( U = \sqmatrix{u_{00}}{u_{01}}{u_{10}}{u_{11}} \), and integer $m \in
\setof{0,1,2,\ldots}$, the operator $\wedge_m(U)$ which acts on $m+1$ qubits,
is defined as
\[
\wedge_m(U) \ket{x_1,\ldots,x_m,y}  =  
      \left\{
         \begin{array}{l}
            u_{y0}\ket{x_1,\ldots,x_m,0} + u_{y1}\ket{x_1,\ldots,x_m,1}
               \mbox{ if } \wedge^m_{k=1} x_k = 1 \\
            \ket{x_1,\ldots,x_m,y} \mbox{ if } \wedge^m_{k=1} x_k = 0
         \end{array}
      \right.
\]
for all $x_1,\ldots,x_m, y \in \setof{0,1}$, where ``$\wedge$'' denotes logical
`and.'

In other words, $\wedge_m(U)$ is an $m$-fold control-$U$ gate.
For example, if
\(
   U = X = \sqmatrix{0}{1}{1}{0},
\)
then $\wedge_1(U)$ is the CNOT gate, and $\wedge_2(U)$ is the Toffoli gate.
More detailed discussion about this notation can be found in~\cite{elem-gate}.

\begin{figure}[h]
   \includegraphics{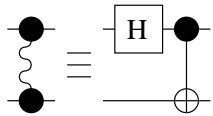}
   \caption{\label{fig:entan-gate} This diagram represents an entangling gate,
      denoted by $E$. This gate can be locally implemented using a Hadamard
      gate and a CNOT gate, as shown. A distributed version of the entangling
      gate is discussed in section \ref{sec:impt-em}.  }
\end{figure}

To reduce the complexity of diagrams, we introduce a diagram for an entangling
gate, called an $E$ gate, as shown in figure~\ref{fig:entan-gate}.  Two qubits
are entangled by this $E$ gate. In particular, if the input state is
$\ket{00}$, the output state will be $\onesqrt(\ket{00}+\ket{11})$.

\begin{figure}
   \includegraphics{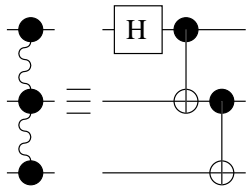}
   \caption{\label{fig:ghz-gate} This diagram represents a generalized
      entangling gate for 3 qubits, denoted by $E_3$. It creates the
      GHZ state if the input is the state $\ket{000}$. This gate can be
      locally implemented by a Hadamard gate and two CNOT gates, as shown.}
\end{figure}

The entangling gate $E$ can be generalized to an $m$-fold entangling gate,
denoted by $E_m$, which creates the cat state $\onesqrt(\ket{00\ldots 0} +
\ket{11\ldots 1})$ upon the input state $\ket{00\ldots 0}$. The $E_3$ gate is
illustrated in figure~\ref{fig:ghz-gate}.

\begin{figure}
   \includegraphics{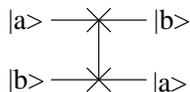}
   \caption{\label{fig:swap-gate} A swap gate. The discussion of distributed
       implementation of this gate can be found on section~\ref{sec:dist-swap}
   }
\end{figure}

Finally, figure~\ref{fig:swap-gate} shows a diagram for the swap gate.
Section~\ref{sec:impt-em} discusses the distributed implementations of the
$E_m$ gate and the swap gate.

\section{\label{sec:dist-ctrl} Distributing Control via a Cat-like State}

In this section, we discuss how a cat-like state can distribute control over
multiple computers.  {\em This is the key idea of the construction of non-local
interaction presented in this paper.} We will demonstrate this idea using the
simplest control gate, i.e.,  the CNOT gate.

The CNOT gate $\wedge_1(X)$ is defined on a two-qubit system as follows: For
any control qubit $\alpha\ket{0}+\beta\ket{1}$ and target bit $\ket{t}$,
\begin{eqnarray}
   \wedge_1(X)((\alpha\ket{0}+\beta\ket{1})\ket{t})
   & = & \alpha\ket{0}\ket{t} + \beta\ket{1} X(\ket{t}) \label{eqn:samp_ctrl}.
\end{eqnarray}
We assume that a cat-state $\onesqrt(\ket{0\ldots 0}+\ket{1\ldots 1})$ has
already been shared between multiple computers. Then cat-entangler can be used
to transform a control qubit and a cat-state into a cat-like state. 
As a result, equation~\ref{eqn:samp_ctrl} becomes 
\begin{equation} 
   \alpha\ket{00\ldots 0}\ket{t} + \beta\ket{11\ldots 1}X(\ket{t}),
   \label{eqn:multi_ctrl}
\end{equation}
which can be rewritten as
\begin{equation}
   \wedge_1(X)((\alpha\ket{00\ldots 0}+\beta\ket{11\ldots 1})\ket{t}).
   \label{eqn:dist_ctrl}
\end{equation}
  
\begin{figure}[h]
   \includegraphics{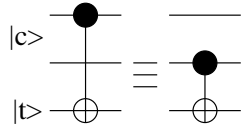}
   \caption{\label{fig:eqv-ctrl} Since line one and line two are in the state 
      $\ket{c}=\alpha\ket{00}+\beta\ket{11}$, the target qubit $\ket{t}$
      can be controlled by either line one or two.
   }
\end{figure}

Equations~\ref{eqn:multi_ctrl} and~\ref{eqn:dist_ctrl} show that we can use
any qubit, shared in the cat-like state, as a control line. For example, the
two circuits in figure~\ref{fig:eqv-ctrl} have an equivalent effect on the
target bit, assuming that the state of lines one and two is the cat-like
state $\alpha\ket{00}+\beta\ket{11}$. 
 
\begin{figure}[h]
   \includegraphics{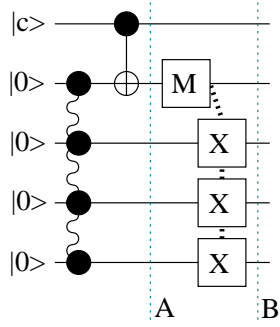}
   \caption{\label{fig:create-ctrl} Given a qubit $\ket{c}=\alpha\ket{0}+\beta
      \ket{1}$ and a generalized GHZ state $\onesqrt(\ket{0000} + \beta
      \ket{1111})$, created by an entangling gate $E_4$, one can create a
      cat-like state $\alpha\ket{0000}+\beta\ket{1111}$ with the above circuit.
   }
\end{figure}

A cat-entangler is shown in figure~\ref{fig:create-ctrl}.  The box
labeled by $M$ is a standard basis measurement. Since a result of the
measurement (represented by a dotted line) is a classical bit, we can
distribute and reuse the result to control many $X$-gates at the same
time.

\begin{theorem}
   \label{thm:dist-ctrl}
   Given a qubit $\ket{c}=\alpha\ket{0}+\beta\ket{1}$ and an $m$-fold cat state
   $\onesqrt(\ket{00\ldots 0_m}+\ket{11\ldots 1_m})$, a cat-like state
   $\ket{\psi_c}=\alpha\ket{00\ldots 0_m}+\beta\ket{11\ldots 1_m}$ can be
   created by a CNOT gate, local operations (i.e., one measurement and
   $X$-gates), and classical communication.
\end{theorem}

Proof:
   A quantum circuit that creates an $m$-fold cat-like state can be
   generalized from the circuit shown in figure \ref{fig:create-ctrl}.  Assume
   that an m-fold GHZ state is created by a $E_m$ gate. 
 
At point A, after applying the CNOT gate, the state of the circuit is 
\begin{eqnarray}
  \onesqrt (\alpha\ket{000\dots 0_m}+\alpha\ket{011\ldots 1_m} 
  +\beta\ket{110\ldots 0_m}+\beta\ket{101\ldots 1_m} )
\end{eqnarray}
%
%
After the measurement, the state is either
\begin{subequations}
   \begin{equation}
   \alpha\ket{0\underline{0}0\dots 0_m}+\beta\ket{1\underline{0}1\ldots 1_m}
   \end{equation}
   or
   \begin{equation}
   \alpha\ket{0\underline{1}1\dots 1_m}+\beta\ket{1\underline{1}0\ldots 0_m},
   \end{equation}
\end{subequations}
where the underlined qubit is the measured qubit. 

Assume that the result of measurement is a classical bit $r$. After applying
$X$ controlled by the classical bit $r$ (represented by a dotted line on the
$X$ gates), the state at point B is 
\begin{eqnarray}
   \alpha\ket{0\underline{r}0\dots 0_m}+\beta\ket{1\underline{r}1\ldots 1_m}.
\end{eqnarray}
Hence, an m-fold cat-like state shared between the qubit $\ket{c}$ and
other qubits from the cat state (except the measured qubit) is created.

\begin{figure}
   \includegraphics{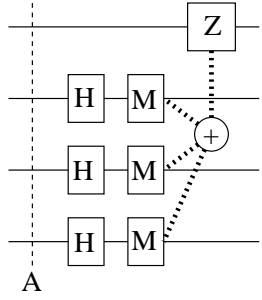}
   \caption{\label{fig:restore-ctrl} 
      An input state at point $A$, $\alpha\ket{0000}+\beta\ket{1111}$, is
      transformed by a cat-disentangler into a state
      $(\alpha\ket{0}+\beta\ket{1})\ket{r_2r_3r_4}$, where $r_k$ is the result
      of measurements on line $k$. The phase-flip gate ($Z$-gate) is controlled
      by the mod 2 sum of the $r_k$, written by $\oplus_k r_k$.} 
\end{figure}

To complete the non-local CNOT operation, the control line must be disentangled
and restored. This can be accomplished by a cat-disentangler, as demonstrated
in figure~\ref{fig:restore-ctrl}.
 
\begin{theorem}
   A state $\ket{\psi_c} = \alpha\ket{00\ldots 0_m} +
   \beta\ket{11\ldots 1_m}$ can be transformed into a state $\alpha\ket{0} +
   \beta\ket{1})\ket{r}$, where $r \in \setof{0,1,\ldots,2^{m-1}-1}$ is
   the result of $m$ $1$-qubit measurements, by a cat-disentangler, which can be
   generalized from the circuit shown in figure~\ref{fig:restore-ctrl}.
\end{theorem}
Proof:
   Assume the input of the circuit is $\ket{\psi_c}$. After applying Hadamard
   transformations, the state of the circuit is 
   \begin{eqnarray}
      \alpha\ket{0}\sum_{r=0}^{2^{m-1}-1} \ket{r} +
      \beta\ket{1} \sum_{r=0}^{2^{m-1}-1} (-1)^{\oplus_k r_k} \ket{r}
   \end{eqnarray}
   where the binary representation of $r$ is $r_1r_2\cdots r_{m-1}$.

   After the measurements, the state becomes
   \begin{eqnarray}
      (\alpha\ket{0} + (-1)^{\oplus_k r_k}\beta\ket{1}) \ket{r}
   \end{eqnarray}
   where $r_i$ is the result of the measurement on line $i+1$.
   
   To correct the phase, we use the result of the computation $\oplus_k r_k$
   to control the $Z$ gate applied to the first line. Hence, the state at the
   end of the circuit is $(\alpha\ket{0}+\beta\ket{1})\ket{r}$.

By considering the circuit in figure~\ref{fig:restore-ctrl}, the first line can
be switched with another line.  Therefore, the control state can be restored to
any qubit involved in the cat-like state. Furthermore, we can leave more qubits
to be untouched, which means we do not apply Hadamard and measurement on one or
more qubits. As a result, the remaining qubits form a smaller cat-like state.

\subsection{Constructing a Teleportation Circuit}

\begin{figure}[h]
   \includegraphics{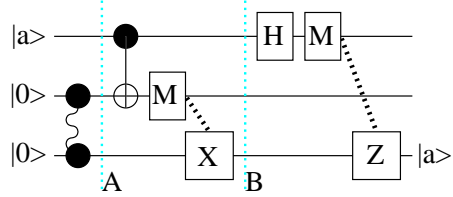}
   \caption{\label{fig:teleport} This figure shows a construction of a
      teleportation circuit using the cat-entangler and cat-disentangler. An
      unknown quantum state $\ket{a}$ is teleported from the first line to the
      third line.}
\end{figure}

A teleportation circuit can be considered as a composition of the cat-entangler
and cat-disentangler as shown in figure~\ref{fig:teleport}.

If the cat-entangler is applied, then the state of qubits $1$ and $3$ is in the
cat-like state $\alpha\ket{00}+\beta\ket{11}$. We next apply the
cat-disentangler. However, the unknown qubit $\ket{a}$ is restored on line $3$,
not on line $1$. 

After carefully considering the teleportation circuit, the group of operations
applied on the first two qubits after the entangling gate, is actually the Bell
basis transformation followed by standard basis measurement. These operations
are equivalent to a complete Bell measurement.  Furthermore, the result of the
measurement is used to control the $X$ and $Z$ gates, as in the teleportation
circuit described in~\cite{teleport}.

\section{\label{sec:ctrl-gates} Constructing Non-local Control Gates}

In this section, we begin by discussing the construction of a non-local CNOT
gate. Then we show how to construct efficient distributed control gates in
different situations. We assume that we have distributed entangling gates,
which create and share a cat state among multiple computers. Construction of
distributed entangling gates is described in section~\ref{sec:impt-em}.

\subsection{\label{sec:ctrl-not} Constructing a Non-local CNOT Gate}

\begin{figure}
   \includegraphics{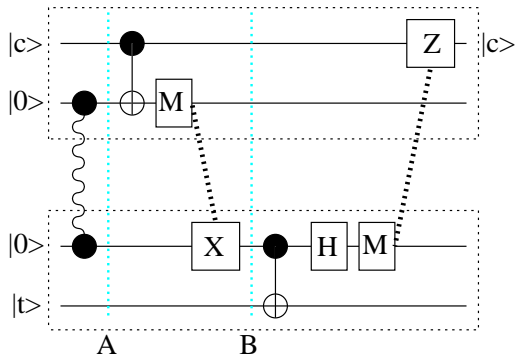}
   \caption{\label{fig:nonlocal-cnot}
      After applying the cat-entangler, the state of the control
      line and the third line becomes a cat-like state $\alpha\ket{00} +
      \beta\ket{11}$. Instead of using the control line (the first line)
      directly, line $3$ can be used to control line 4. Then we apply a
      cat-disentangler to restore the control qubit. Each computer
      boundary is indicated by a tightly dotted box.
   }
\end{figure}

By observing that a control line can be distributed by a cat-like state, a
non-local CNOT gate can be implemented as shown in
figure~\ref{fig:nonlocal-cnot}.

\begin{theorem}
   \label{thm:cnot-proof}
   Given a control line $\ket{c}=\alpha\ket{0}+\beta\ket{1}$ and an entangled
   pair $\epr$ created by an entangling gate, a non-local CNOT gate can be
   implemented from the cat-entangler and cat-disentangler, as shown in the
   figure~\ref{fig:nonlocal-cnot}.  
\end{theorem}
Proof: After applying the cat-entangler, the state at point $B$ is
\begin{eqnarray}
   (\alpha\ket{0r0} +  \beta\ket{1r1})\ket{t},
\end{eqnarray}
where $r$ denotes the result of the measurement.

Since line 1 and line 3 are in a cat-like state $\alpha\ket{00} +
\beta\ket{11}$. We can use either line 1 or line 3 to control the $X$ gate.
In this case, we use line 3, which is on the local machine.

Because the control line does not change after applying CNOT, the state of line
1 and line 3 remains in a cat-like state. Therefore after applying the
cat-disentangler, the control qubit is restored to line 1.
Hence, we have completed a non-local CNOT operation. 

This circuit is proven to be the optimal implementation by~\cite{non-local,
non-local-content}, i.e. one ebit and two cbits are necessary and sufficient
for implementing a non-local CNOT gate. Furthermore, the cat-entangler and
cat-disentangler can be applied to create a $\wedge_1(U)$. In many
cases, we have an efficient implementation of $\wedge_1(U)$.

\subsection{\label{sec:small-cu} Constructing of Small Non-local Control Gates}

In this section, we assume that a unitary transformation $U$ is small enough to
implement on one computer, but that the control line is on a different
computer. Moreover, the transformation $U$ is composed of a number of basic
gates, i.e, $U = U_1\cdot U_2\cdots U_k$, for some integer $k$.

\begin{figure} 
   \includegraphics{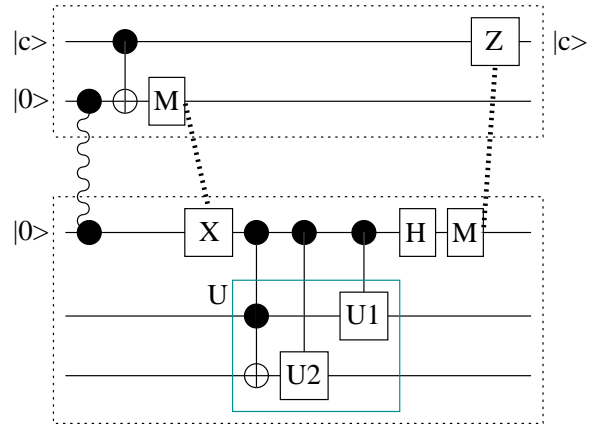}
   \caption{\label{fig:nonlocal-cu}
      Assume $U = U_1\cdot U_2\cdot CNOT$.  Then $\wedge_1(U)=\wedge_1(U_1)
      \cdot \wedge_1(U_2) \cdot \wedge_1(CNOT)$ can be distributed as shown.
      The control line needs to be distributed only once, because it can
      be reused. This implementation allows the cost of distributing the
      control qubit to be shared among the elementary gates. 
   } 
\end{figure}

After a cat-entangler creates a cat-like state, this establish distributed
control between two computers. Since the control line can be reused, only one
cat-entangler is needed. Therefore, one ebit and two cbits are needed to
implement this type of non-local $\wedge_1(U)$ gate. This idea is demonstrated
in figure~\ref{fig:nonlocal-cu}, where $U = U_1\cdot U_2\cdot CNOT$.

Because the control line is reused, the cost of implementing the
$\wedge_1(U_i)$ gates can be shared among the basic gates. In other words, each
non-local $\wedge_1(U_i)$ can be implemented using only $\oneof{k}$ ebits and
$2 \over k$ cbits, asymptotically.

\subsection{\label{sec:large-cu} Constructing Large Non-local Control-U Gates}

In this section, we assume that a unitary transformation $U$ is too big to
implement on one computer. Therefore, $U$ has to be decomposed into a number of
smaller transformations, where each transformation can be implemented on one
computer. In this setting, there are two scenarios to consider.

\begin{figure}
   \includegraphics{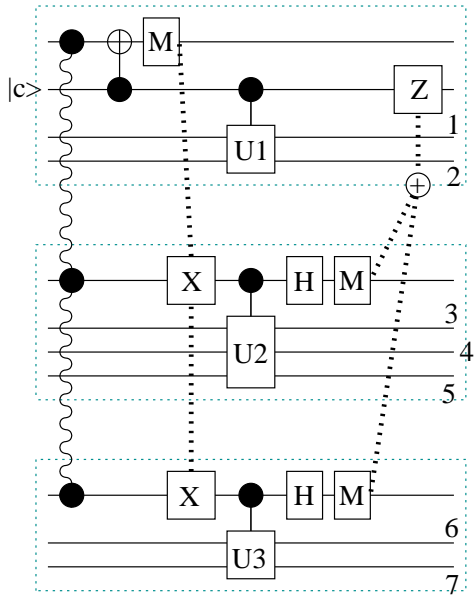}
   \caption{\label{fig:exm-mulparty} In this case, the transformation $U$ can
   be decomposed into a set of small transformations. The cat-like state allows
   the control to be distributed among multiple computers, and also allows each
   computer to execute in parallel.} 
\end{figure}

In the first scenario, we assume that there are no shared qubits between the
components of $U$, i.e. $U$ is decomposed into smaller transformations, where
each transformation is applied to different qubits.  For example, assume $U$ is
a unitary transformation for a 7-qubit system, and $U = U_1\cdot U_2\cdot U_3$,
where $U_1$ is a unitary transformation acting on qubits 1 and 2, $U_2$ is a
unitary transformation acting on qubits 3, 4, and 5, and $U_3$ is a unitary
transformation acting on qubits 6 and 7. In this case, not only the cat-like
state can be used to distribute a control line among three computers, but 
$\wedge_1(U)$ can also be executed in parallel, as demonstrated in
figure~\ref{fig:exm-mulparty}.
 
In the second scenario, there are shared qubits between these transformations.
For example, assume a transformation $U$ is a transformation on a 3-qubit
system, and $U = U_1\cdot U_2\cdot U_3\cdot U_4$, where $U_1, U_4$ acts on
qubit 1 and 2 and $U_2, U_3$ acts on qubit 2 and 3.

\begin{figure}[h]
   \includegraphics{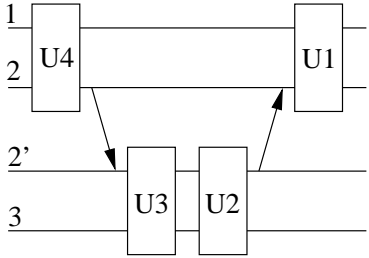}
   \caption{\label{fig:share-one} $U$ can be decomposed into a set of smaller
      transformations that can be implemented on one computer. However, they
      share qubit number $2$. A distributed version of this case can be
      implemented by transporting or teleporting qubit number $2$ back and forth
      between the top and the bottom machine. Line $2$ becomes line $2'$ in the
      bottom computer.}
\end{figure}

In this case, the shared qubits can be physically transported or teleported
from one computer to another. However, sending this shared qubit introduces a
communication overhead, which is not present in a non-distributed
implementation of $\wedge_1(U)$.


\subsection{\label{sec:mctrl} Constructing a $\wedge_m(U)$ Gate}
 
\begin{figure}
   \includegraphics{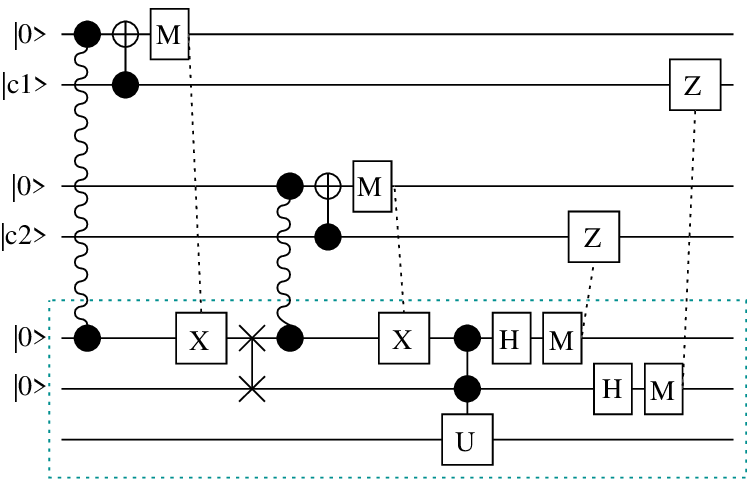}
   \caption{\label{fig:mctrl-2u} Both control lines $\ket{c_1}$ and $\ket{c_2}$
      are distributed via two cat-like states to the bottom machine so that
      $\wedge_2(U)$ can be implemented locally. The swap gate moves the
      control line from the channel qubit and resets it to $\ket{0}$, so that
      the channel qubit can be reused.}
\end{figure}

A non-local $\wedge_m(U)$ can simply be implemented by using $m$ cat-like pairs
to distribute $m$ control qubits to one machine, and then to implement
$\wedge_m(U)$ locally, as suggested in~\cite{non-local}.
Figure~\ref{fig:mctrl-2u} shows an instance of $\wedge_2(U)$.

However, this requires the computer to have enough qubits to implement
$\wedge_m(U)$ locally. Therefore, the number of non-local control lines is
limited by the number of qubits available on one computer.

\begin{figure}
   \includegraphics{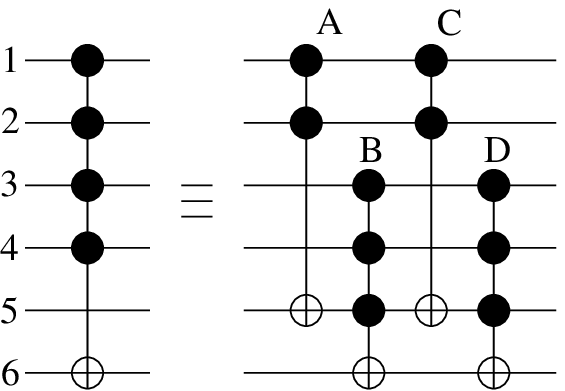}
   \caption{\label{fig:mctrl-lin} In this case, $m=6$ and $n=2$. The
      $\wedge_4(X)$ gate is simulated by two $\wedge_2$ gates and
      two $\wedge_3$ gates as shown.}
\end{figure}
 
Barenco et al show that the $\wedge_{m-2}(X)$ gate can be linearly simulated on
$m$-qubit system~\cite{elem-gate}. For example, for any $m \ge 5$, $2 \le n
\le m-3$, a $\wedge_{m-2}(X)$ gate can be implemented as two $\wedge_n(X)$
gates and two $\wedge_{m-n-1}(X)$ gates on an $m$-qubit system.  An
implementation for the case of $m=6$ and $n=2$ is shown in figure
\ref{fig:mctrl-lin}. 

This technique can be used to break down a large $\wedge_m(U)$ gate into a
sequence of smaller gates. After thus making the number of control lines small
enough, we can distribute these control lines to one machine, and implement
each gate locally.

\begin{figure}
   \includegraphics{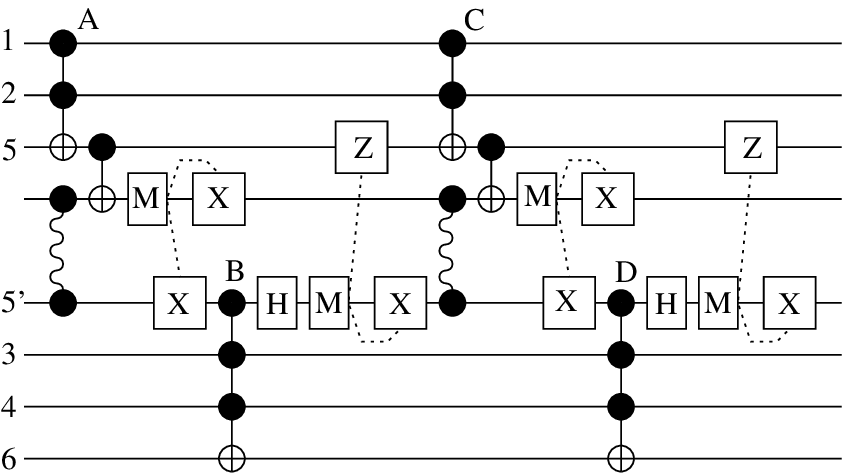}
   \caption{\label{fig:mctrl-lin-dist} This circuit presents a
      distributed version of $\wedge_4(X)$, where $m=6$ and $n=2$.
      Line 5 becomes line 5' on the bottom machine. }
\end{figure}

A distributed implementation of figure~\ref{fig:mctrl-lin} is demonstrated in
figure~\ref{fig:mctrl-lin-dist}.  We group lines 1, 2, and 5 together on the
top machine. Then we distribute line 5 to line 5' and then use it to perform a
control operation with lines 3 and 4 applied to line 6 in the bottom machine.
 
\section{\label{sec:impt-em} Establishing and Refreshing an Entangled Pair}
Before any non-local CNOT and teleportation operations are performed, an
entangled pair must be established between channel computers. Moreover, after
the operation is finished, the channel qubits must be reset to refresh the
pair. 
We discuss establishing and refreshing procedure in the next section via the
construction of entangling gates.

\subsection{\label{sec:ent-e2} Constructing the $E_2$ Gate}

Because establishing entanglement can not be accomplished by any local
operations, channel qubits must be on one machine. The simplest way to
establish an entangled pair is described as follow:

To establish an ebit between a remote and a local computers, the remote
computer sends a channel qubit to the local machine. Then, the local machine
entangles the remote channel qubit with one of the local channel qubits. Then
one qubit of the pair is sent back to the remote machine. This process
establishes one ebit by sending two qubits, one in each direction.

\begin{figure}
   \includegraphics{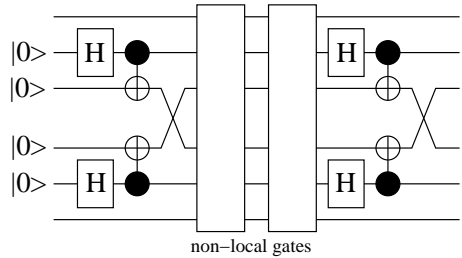}
   \caption{\label{fig:ent-e2} Assuming each machine holds a pair of channel
   qubits. Each machine entangles its channel pair, and then exchanges one
   qubit of the pair with another machine.} 
\end{figure}

If we allow each machine to have two channel qubits, then two ebits can be
established by sending two qubits.  To do so, each computer entangles its own
channel qubits, then exchanges one qubit of the pair with the other computer,
as demonstrated in figure~\ref{fig:ent-e2}. As a result, one ebit is
established with the cost of sending one qubit, asymptotically. To refresh the
entanglement, the procedure is simply repeated. However, the channel qubits
must be reset to state $\ket{0}$ before they can be re-entangled.

\begin{figure}
   \includegraphics{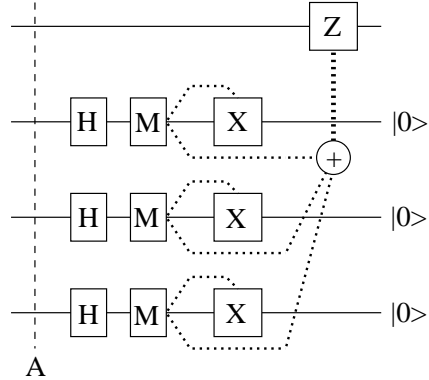}
   \caption{\label{fig:reverse-reset} 
      This figure demonstrates how the channel qubits can be reset to
      $\ket{0}$ by considering the result of the measurement. Assume the input
      state at point A is $\alpha \ket{0000} + \beta \ket{1111}$. The output
      of the circuit is the state $(\alpha \ket{0} + \beta \ket{1})
      \ket{000}$.
   } 
\end{figure}

Channel qubits need to be in the state $\ket{0}$ before being entangled.
Therefore the channel qubits must be reset to the state $\ket{0}$ before the
entangled pair can be refreshed. Because the measurement result reveals the
current state of the measured qubit, we can use the measurement result to
control the $X$-gate to set the channel qubit to $\ket{0}$, as shown in
figure~\ref{fig:reverse-reset}.

\subsection{\label{sec:ent-em} Constructing the $E_m$ Gate}

\begin{figure}
   \includegraphics{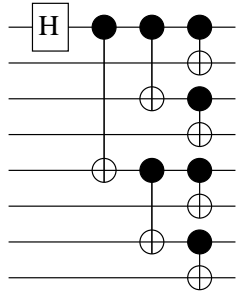}
   \caption{\label{fig:em-gate} This figure represent an efficient
   implementation of $G_8$, using a Hadamard gate and 7 CNOT gates. A
   generalized GHZ state can be created after applying $O(\log  m)$ gates.
   However, this implementation requires more channel qubits.} 
\end{figure}

A non-distributed version of $E_m$ gate can be linearly implemented using a
Hadamard gate, and a sequence of CNOT gates as shown in figure
\ref{fig:ghz-gate} (for three qubits case.) However, it takes $O(m)$ steps to
create an m-fold GHZ state.  An efficient way of implementing a
non-distributed version of an $E_m$ gate is shown in figure~\ref{fig:em-gate}.
This implementation utilizes a binary tree idea to reduce the number of steps
from $O(m)$ to $O(\log m)$.

A distributed version of $E_m$ gate can be implemented by using non-local CNOT
gates. In this process, we exchange $m-1$ ebits with the $m$-fold cat state by
replacing CNOT gates with non-local CNOT gates. 

In the linear implementation of $E_m$, each computer requires at least two
channel qubits to establish entanglement with its neighbors. This number is
fixed regardless of the number of computers involved.

On the contrary, in the binary implementation, the number of channel qubits on
each computer increases with respect to $m$. In fact, the number of channel
qubits required is $O(\log m)$.  This is true because the first computer has to
simultaneously establish $O(\log m)$ entangled pairs with others computers, as
shown in figure~\ref{fig:em-gate}, before an $m$-fold cat state is created.
However, the trade off is justified because the number of steps required to
establish the $m$-fold cat state is reduced from $O(m)$ to $O(\log m)$.

%
%

\subsection{\label{sec:recont-tel} Re-constructing a Teleportation Circuit}

\begin{figure}[h]
   \includegraphics{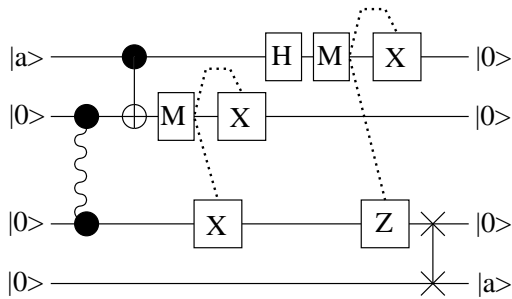}
   \caption{\label{fig:reset-tel} The remote channel qubit can be reset by
      swapping with a empty qubit ($\ket{0}$). In addition, after
      teleportation, the original qubit is reset to $\ket{0}$ state.
   }
\end{figure}

In the implementation of a non-local CNOT gate, measurements are performed on
both channel qubits.  Therefore, the channel qubits can be reset by considering
the measurement results.  In the teleportation circuit, the remote channel
qubit is not measured.  Moreover, it holds the unknown state which is
teleported by the circuit.  Pati and Braunstein show in~\cite{nodelete} that an
unknown qubit can not be deleted. To solve this problem, we can swap the remote
channel qubit with $\ket{0}$ held by another qubit in the register on the same
machine.  By swapping, the unknown qubit is preserved and the channel qubit is
reset.  The construction of the teleportation circuit, with channel qubits
reset, is shown in figure~\ref{fig:reset-tel}.

One remaining question is how many $\ket{0}'s$ (or empty qubits) are needed to
support the teleportation protocol. We observe that, after teleportation, the
original unknown qubit on the source machine is reset to state $\ket{0}$.
Therefore, we can use this qubit as an empty qubit in the next inbound
teleportation. However, the required number of empty qubits still depends on
how many qubits are teleported into the machine before another qubit is
teleported out.  In other words, we can consider the $\ket{0}$ qubit as an
empty space. To teleport a qubit in, we need an empty space to receive the
qubit. The more qubits teleported into the machine, the more spaces needed.

\subsection{\label{sec:dist-swap} Constructing a Distributed Swap Gate}

A distributed swap gate can be implemented by using two teleportations to
exchange two qubits. If each machine has two channel qubits involved in a
swapping operation, we do not need any empty qubits in the registers. The
channel qubit becomes a swapping buffer, because the first teleportation
creates an empty qubit for the second teleportation to use. However, if we have
only one channel qubit in a swapping operation, one empty qubit is needed for a
swapping buffer.

In~\cite{non-local-content}, Collins, Linden, and Popescu show that an optimal
non-local swap gate requires two ebits and two cbits in each direction.
Therefore, implementing a swap gate by two teleportations is an optimal
implementation.

\section{\label{sec:fourier} Distributed Fourier Transform}
\begin{figure*}
   \includegraphics{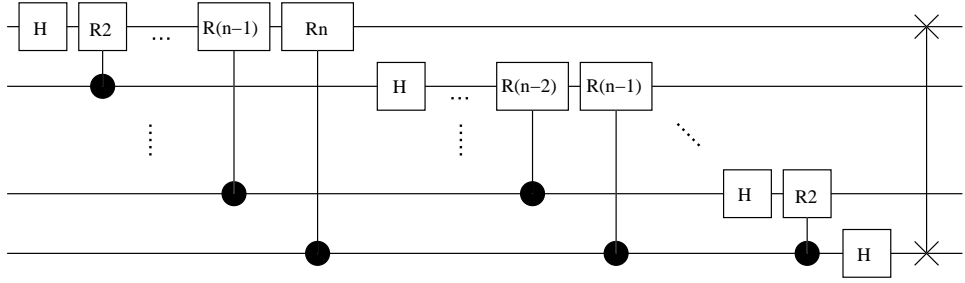}
   \caption{\label{fig:fourier} The Fourier transformation is a sequence
      of Hadamard and control-$R_k$ gates. At the end, a swapping gate
      is used to reset the order of the qubits.}
\end{figure*}
\begin{figure*}
   \includegraphics{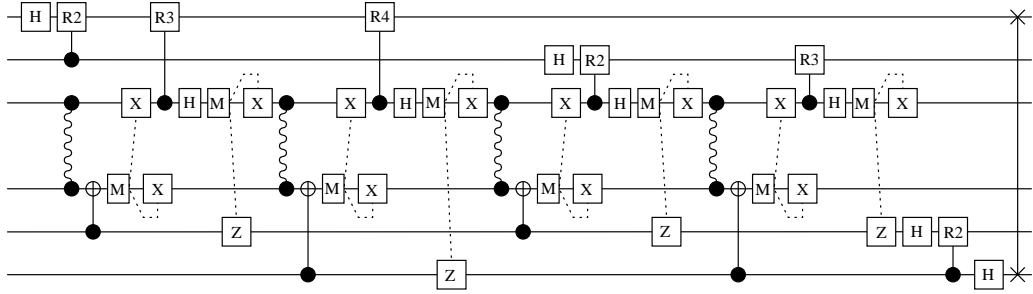}
   \caption{\label{fig:dist-fourier} By applying non-local control $R_k$
      gates, we construct a distributed Fourier transformation. This figure
      shows the distributed Fourier transformation for 4 qubits which is
      implemented on two machines. The swap gate can be implemented by
      teleporting qubits back and forth between two computers.} 
\end{figure*}
The quantum Fourier transform is a unitary transformation defined on
basis states as follows,
\begin{equation}
   \ket{j} \rightarrow \oneof{\sqrt{2^n}} 
                       \sum_{k=0}^{2^n-1} e^{2\pi \imath jk/2^n}\ket{k},
\end{equation}
where $n$ is the number of qubits. See~\cite{factoring,nc-qcqi} for more
details.

Shor presents the quantum Fourier transformation in~\cite{factoring} as a
routine used in the factoring algorithm. Later on, it becomes a standard
routine used in many quantum algorithms, such as phase estimation, and the
hidden subgroup algorithms. An efficient circuit for the quantum Fourier
transformation, found in~\cite{nc-qcqi}, is shown in figure~\ref{fig:fourier}.
The gate $R_k$ is defined as follows,
\begin{eqnarray} 
	R_k = \sqmatrix{1}{0}{0}{e^{2\pi \imath/2^k}},
\end{eqnarray}
where $k \in \setof{2,3,\ldots}$.

We use the construction of $\wedge_1(U)$ to implement a distributed version of
the Fourier transformation. The swap gate can be implemented by teleporting
qubits back and forth between two computers. 

The communication resources needed depend on how many non-local control gates
have to be implemented. Assume that we implement the Fourier transformation on
$m$ computers, and each computer has $k$ qubits, not including the channel
qubits. Let $n=mk$. Therefore, there are $(n-1)n/2$ control gates to implement.
Hence, for each computer, there are $(k-1)k/2$ local control gates. The number
of local control gates is $m(k-1)k/2 = (n/m-1)n/2$.  The number of non-local
control gates is $(n-1)n/2-(n/m-1)n/2$ gates.  Hence, the communication
resources are $O(n^2)$. 

An implementation of the distributed Fourier transformation of $4$ qubits is
shown in figure~\ref{fig:dist-fourier}

\section{Conclusion}

The teleportation circuit and the non-local CNOT gate can be implemented using
two primitive operations, cat-entangler and cat-disentangler.  These two
primitive operations work on the idea that a cat-like state can be used to
distribute a control line.  This principle is extended to construct non-local
operations efficiently in many cases.

For example, the communication cost can be shared among elementary gates
because a control line can be reused after being distributed via cat-like
state.  Moreover, the cat-like state allows parallel implementation of some
control gates.

Before using a non-local operation, an entanglement has to be established.
Moreover, it has to be refreshed after being used. Fortunately, by looking at
the measurement results, we can reset channel qubits, and re-establish
entanglement. This procedure works well with non-local CNOT gates. To reset
channel qubits in the teleportation operation, the channel qubits have to be
swapped with empty qubits ($\ket{0}$).

In general, if we can break down a unitary transformation into a sequence of
CNOT gates and one-qubit gates, a distributed version of the unitary
transformation can be simply implemented by replacing CNOT gates with non-local
CNOT gates. Therefore, the communication overhead is dependent on the number of
non-local CNOT gates needed to be implemented. With help from teleportation to
move qubits back and forth between computers, better distributed implementation
can be accomplished.

\end{document}